\documentstyle[aps,preprint]{revtex}

\begin{document}

\title{Sign Change of the Flux Flow Hall Effect in
 HTSC }

\author{ M. V. Feigel'man$^a$,  V. B. Geshkenbein$^
{a,b}$,
A. I. Larkin$^{a,c}$, and V. M. Vinokur$^c$}
\address{$^a$Landau Institute\ for Theoretical Physics, 117940
Moscow, Russia}

\address{$^b$Theoretische Physik, ETH-H\"onggerberg, 8093
Z\"urich, Switzerland}
\address{$^c$Argonne National Laboratory, Argonne, Illinois
60439, USA}

\maketitle
\begin{abstract}

 A novel mechanism for the sign change of
the  Hall effect in the
flux flow region is proposed. The difference  $\delta
n $ between the electron
density at the center of the vortex core and that far
outside the vortex
causes the additional contribution to the Hall
conductivity
$\delta\sigma_{xy}=-\delta n ec/B$.  This contribution
 can be larger than the conventional one in the dirty
case  $\Delta(T)\tau <1$. If the carrier density
inside the core
 exceeds that far  outside,
a double sign change may occur as a function of
temperature.

\end{abstract}

\vspace*{-15.5cm}
\hspace*{12cm}{ \tt ETH-TH/95-9}


\newpage

The sign change of the Hall effect in the mixed state of
 the high
temperature superconductors (HTSC) is
 the most puzzling and controversal phenomenon in the
physics of
magnetic properties of these materials \cite{Hagen}.
In spite of the
numerous attemps to explain this anomaly even the
origin of the sign reversal in the Hall
resistivity  remains unclear
\cite{Hagen}. The sign
change of the Hall
resistivity had also been  observed in the conventional
 superconductors
and thus is not a pecularity of the HTSC \cite{Hagen,con}.
Comparing experimental data for different materials,
Hagen {\it et al.} \cite{Hagen} argued that the sign
change is an
intrinsic property of the vortex motion, and moreover
that the sign
reversal  occurs in the range of
parameters where the transport mean free path $l$
becomes of the order
of the vortex core size $\xi$.

In the present paper we propose an
explanation of the sign reversal in the Hall resistivity .
We show that the sign change may
follow from the difference $\delta n$ between the
 electron density
at the center of the vortex core and the density far
 outside the core.

In order to describe the Hall effect one has to find a
transverse
force experienced by the vortex moving with the
velocity ${\bf v}_L$
under the applied transport current ${\bf j}_T$. There
are two
contributions to the transverse force. The first
one arises from the
nondissipative momentum transfer from the moving
vortex to infinity.
This contribution consists of the Magnus and
Iordanskii \cite{I} forces.
The second contribution stems from the momentum transfer from the
vortex to the normal excitations in the vortex core. The subsequent
absorption of this transfered momentum by the thermal bath due to
scattering of the
normal excitations  leads to  dissipation and the
longitudinal Bardeen - Stephen friction force.

In order to understand both contributions let us derive the dynamic
term in the adiabatic action for a  moving vortex. The effective action
for a superconductor should depend on the phase of the order parameter
 $\chi$ in the following gauge invariant combinations:
$S=S({\bf \nabla}\chi-2e{\bf A}/\hbar c,\partial\chi/\partial t+
2e\phi/\hbar c)$, where ${\bf A}$ and $\phi$ are the vector and scalar
potentials, respectively. By variation of the action with respect to
the phase one gets the current conservation law:
\begin{equation}
{\delta S\over\delta\chi({\bf r},t)}=-{\hbar\over 2e}\nabla{\bf j}+
{\partial\over\partial t}{\delta S\over \delta\partial\chi/\partial t}=0,
\end{equation}
where we use the definition of the  electric current density
${\bf j}=c\delta S/\delta
{\bf A(r},t)$. Because of the continuity equation $\nabla{\bf j}/e+\partial
n/\partial t=0$ ($n$ is the particle density) for the particle
current,
we immediately see that the effective action has to contain the following
topological  term:
\begin{equation}
S_t=-\hbar\int dV dt {n\over 2} {\partial \chi\over\partial t}.
\end{equation}
The factor 1/2  is due to  pairing, and is  absent for the
superfluid Bose system.
This topological  term is irrelevant if $n= const$ and
 $\chi $ is a single valued function
but in the presence of vortices it is just the term in action
 which determines their dynamics.
Let us consider for simplicity the two dimensional case.
Expressing the phase in the presence of a vortex as a sum of a singular
$\Theta({\bf r-R}_L(t))= arg({\bf r}-{\bf R}_L(t))$, (${\bf R}_L(t)$ is the
vortex line position) and a regular contribution,
$\chi=\Theta ({\bf r-R}_L(t))+\chi_r({\bf r},t)$, and taking only
the singular
contribution into account one can rewrite the topological
 term in the action as
\begin{equation}
S_t={\hbar\over 2}\int d^2r dt n\nabla\Theta\dot{\bf R}_L =
\int dt {\bf a(R}_L){\bf v}_L.
\end{equation}
The quantity ${\bf a(R}_L)={\hbar\over 2}\int d^2r n({\bf r-R}_L)
{\bf\nabla}\Theta({\bf r-R}_L)$
has the meaning of the `vector potential' of a fake constant magnetic field.
To see this, one  calculate
\begin{equation}
\nabla\times {\bf a}= {\hbar\over 2}
\oint n\nabla\Theta d{\bf l} =\pi \hbar (n_\infty-n_0),
\end{equation}
where we replaced the surface integral by the two contour integrals at
infinity and around ${\bf r=R}_L$, with
$n_\infty$ and $n_0$ being the particle densities far outside the
vortex
core and on the core axis respectively. This term in action
describes a  particle (the vortex) moving in an `effective magnetic
field' \cite{FL} resulting in the transverse force
\begin{equation}
{\bf F}_\perp={\bf v}_L\times {\bf z}\pi \hbar(n_\infty-n_0),
\end{equation}
analogous to the Lorentz forced experienced by a particle moving in a
magnetic field ({\bf z} is the unit vector along the vortex axis).
Note that this force is independent of charge and is not of
 electromagnetic origin.
For  superfluid Helium  an additional
factor 2 appears with $n$ being the density of the Helium atoms.

For the Galilean invariant case $n_0=0$ , $n_\infty=n_s$, and the force (5)
is just the Magnus force. Based on a similar Berry phase type of arguments
Ao and Thouless \cite{AT} arrived at the conclusion that the relevant
density in Eq.(5) is  always $n_s$ rather than $n$. We believe that
this difference
arises because in their arguments they use a `superfluid wave function'
$\Psi_s^2\propto n_s$, which is ill defined at finite
temperatures
 or in the presence of disorder, where the difference between $n_s$
and $n$ occurs.

In general there are two major differerences
between the (5) and the Magnus force; first, $n_\infty$ is the total density
rather than the superfluid one and thus this part is the sum of the Magnus
and Iordanskii \cite{I} forces. Second, and most important is that
there is
an additional term proportional to the density at the vortex axis.
In our derivation  of the Eq. (4) we have excluded the vortex
axis from the integration, since $n\nabla\Theta$ has a singularity
there if $n_0\neq 0$, and our adiabatic action is
not applicable very close to the core axis.
The fact that
$n_0\neq 0$ means that not all the particles  participate in the
superfluid
motion and there are normal excitations inside the vortex core \cite{CDM}.
As the Magnus force arises from the nondissipative transfer of the momentum
from the vortex to infinity, the other term $-n_0\hbar{\bf v}_L\times {\bf z}$
is due to the transfer of  momentum from the condensate to the normal
excitations inside
the vortex core. This term is just the term obtained by Volovik \cite{V}
who, starting from the BCS theory, derived an effective
action describing the transfer of the momentum from the condensate to
the fermionic modes in the vortex core. The subsequent absorption
of this momentum by the heat bath due to the scattering of these excitations
on the impurities leads to the Bardeen Stephen dissipation.
Thus the hidden assumption in the derivation of the Eq. (5)
was that impurity scattering is so strong that all the momentum
transfered to the normal excitation is absorbed by the heat bath.
 However, for  BCS superconductors we have
$n_\infty-n_0\ll n_\infty$ and the Magnus force is compensated almost
completely \cite{V}. In this case the impurity scattering should be
considered in more details.
Such a calculations have been done long ago
\cite{KK}  without the account for the nonzero $n_\infty-n_0$ difference.
Our goal is to take into account both effects simulteneously and show that
their combination can lead to a change of the overall sign of Hall
conductivity.

 An accurate treatment of  the scattering processes in  the
adiabatic action approach is complicated and is left for future
investigations.
Instead we consider a simple phenomenological model
which is similar
to the original model of Nozi\'eres and Vinen  \cite{NV}, but differs in that
 we take into account both the impurity scattering and the change in
density.
 To this end we consider a  model
of the fully normal core with a carrier density $n_0$
and a sharp boundary at a radius $r_c\simeq\xi$ with
the superconductive material having a density $n$ \cite{BS,NV}. We denote
the
velocity  of the
 normal carriers inside the core  in the laboratory frame  as ${\bf v}_c$ and
look at the transfer of the momentum in the system. The conservation
of the
momentum $d{\bf P}/dt=0$ in the electron  system, with a
 transport current ${\bf j}_T$, and
electric as well as  magnetic fields  present reads
\begin{equation}
{\bf j}_T\times {\bf B}/c+ne{\bf E}-mn_0{\bf v}_cf(B/H_{c2})/ \tau=0
\end{equation}

The first two terms describe the momentum transfer due to the Lorentz-
and the electric field forces. The third term accounts for the
 momentum transfer
due to the impurity scattering ($\tau$ and $m$ are the transport time and the
effective mass respectively). For $B>H_{c2}$, $f(B/H_{c2})=1$, $ne{\bf v}_c=
{\bf j}_T$ and the equation (6) gives the Drude formulas for the longitudinal
and Hall conductivities. For  $B\ll H_{c2}$ the impurity scattering happens
only in the vortex core and $f\propto B/H_{c2}$. If the carrier density
$n_0$ inside the core is equal to that
outside the core $n$, than the transport current ${\bf j}_T$ is equal to
$n_0 e{\bf v}_c$ and we obtain \cite{NV}
$\omega_0\tau {\bf z\times (v}_L-{\bf v}_c)={\bf v}_c$,
where we introduced $\omega_0=eB/mcf(B/H_{c2})$ ($\omega_0\simeq \Delta^2/E_F$
at $B\ll H_{c2}$). The same result can be obtained by writing the steady state
equation for the normal excitation inside the vortex core \cite{grenoble}.
Solving this equation  one finds
 \begin{equation}
{\bf v}_c={\omega_0\tau\over 1+(\omega_0\tau)^2} {\bf
z\times v}_L+{(\omega_0\tau)^2\over
1+(\omega_0\tau)^2}{\bf v}_L,
\end{equation}
which coincides with the result of the microscopic calculations
in the relaxation time approximation \cite{KK}.
If, however, $n_0\not= n$ then $n_0 e{\bf v}_c$ can not
be equal to the ${\bf j}_T$.
 In the reference frame moving with the vortex the current conservation gives
  $n_0 e{\bf\tilde v}_c={\bf\tilde j}_T $. Going to the laboratory
frame we have $n_0 e{\bf v}_c={\bf j}_T+\delta n e {\bf v}_L$,
where $\delta n=n_0-n_{\infty}$
Inserting in  Eq.(6) one sees that the equation for ${\bf v}_c$
(7) remains unchanged and
\begin{equation}
{\bf j}_T={en_0\omega_0\tau{\bf
z\times v}_L\over 1+(\omega_0\tau)^2}+e
({n_0(\omega_0\tau)^2\over
1+(\omega_0\tau)^2}-\delta n)
{\bf v}_L.
\end{equation}
where the first term in r. h. s.
is the Bardeen Stephen longitudinal conductivity
and  the second
term is the  Hall conductivity.
The $\delta n$ term rewritten as the transverse force is just the
term which we derived in Eq. 5 considering the adiabatic action.
{}From these topological arguments (see also \cite{V}) it follows
that $\delta n$ in Eqs. (8) is not some averaged change in
density but is the difference between the electron density
on the axis of the vortex core  and that far outside the core.
 The  transverse force  can be rewritten as
$ \pi\hbar(n-{n_0\over
1+(\omega_0\tau)^2}){\bf v}_L\times {\bf z}$,
where the first term is  the Magnus  force and the second one is the
force due to  impurity scattering which cancells the Magnus
force almost completly in conventional situation \cite{KK,V}.

{}From the Eq. (8) we obtain the  Hall
 conductivity
\begin{equation}
\sigma_{xy}={n_0ec\over B}{(\omega_0\tau)^2\over
1+(\omega_0\tau)^2}-{\delta n ec\over B}
\end{equation}
The additional contribution to the flux flow Hall conductivity
$\delta\sigma_{xy}=-\delta n ec/B$ is  our  main result.

The above considerations are  valid for a model uncharged
superconductor, and in that case $\delta n/n \sim (\Delta/E_F)^2$ \cite{V}.
In real superconductors the Coulomb screening is always present, and
suppresses strongly any inhomogenities in the charge density
distribution, and  the total
charge of the vortex becomes zero. We will see, however, that the
screening has no effect on the value of the Hall conductivity, though the
latter can't be expressed any more in terms of the density difference $\delta
n$. In order to account for the screening effect we supplement the
superconductive Lagrangian ${\cal L}_{sc}$ by the Coulomb terms: ${\cal
L}_{tot} = {\cal L}_{sc} + {\cal L}_C$, where
\begin{equation}
{\cal L}_C = \frac{1}{8\pi}\sum_{q}
 {\bf E}_{\parallel}({\bf q}){\bf E}_{\parallel}({\bf -q})
\varepsilon({\bf q})
\end{equation}
where ${\bf E}_{\parallel}({\bf q})$ is the Fourier-component of the
longitudinal electric field, ${\bf E}_{\parallel}({\bf r}) = -\nabla \phi$,
and $\varepsilon({\bf q})$ is the dielectric function,
whose $q \rightarrow 0$ limit determines the screening length $r_D$,
$\varepsilon(q) =1+1/ r_D^2 q^2$; normally $r_D \ll \xi_0$.
A distribution of the electric
potential around the vortex is determined now by the equation
 $\delta {\cal L}/\delta \phi (r) =-e \delta\tilde n(r)  +
 (\phi/r_D^2 - \nabla^2 \phi)/4\pi = 0$, whereas the charge density
$e\delta n =- \nabla^2 \phi/(4\pi) = e\delta \tilde n - \phi/(4\pi r_D^2)$,
and we introduced for future convenience the notation $e\delta \tilde n(r) =
-\delta {\cal L}_{sc}/\delta \phi (r)$.
In the weak screening limit $r_D \gg \xi$
one would get $\delta n(r) \approx \delta\tilde n(r)$, whereas at $r_D \ll
\xi$ the density is almost constant and $\delta n(r) \simeq \delta \tilde n(r)
(r_D/\xi)^2 \sim n_0 (\Delta/E_F)^4$ \cite{Schmid}, whereas $\phi(r) = 4\pi
r_D^2
\delta\tilde n(r)$. The key point now is that
 the Hall conductivity can be expressed via the value of $\delta\tilde n =
\delta\tilde n(0)$ which {\it does not depend on screening}. It follows from
the fact that the Coulomb part of the Lagrangian ${\cal L}_C$  depends
upon the longitudinal electric field only, and thus does not contain
contribution from the singular vortex-induced phase $\chi_v({\bf
  r}) = \Theta({\bf r}-{\bf r}_L)$.
 Therefore the `topological' contribution to
the Hall conductivity is determined by the effective action $S_t$, Eq.(2),
with $n(r)$ replaced by $\delta\tilde n(r)$, so the result, Eq.(9), is
recovered up to the replacement of $\delta n$ by $\delta\tilde n=
\partial \Omega_{sc}/\partial \mu$ ( where $\mu$ is a chemical potential).
 In  leading order of $(Tc-T)$ $\delta\tilde n$ can be
 expressed via the experimentally accessible quantities:
\begin{equation}
\delta\tilde n =- \frac{H_c^2(T)}{4\pi}\frac{\partial ln (T_c - T)}
{\partial \mu}
\end{equation}

We will show now that the  $\delta n$ term in the conductvity we found
is just the term which was obtained in Time Dependent Ginzburg Landau
(TDGL) model \cite{D,IKK}. We will follow the approach developed in
 \cite{AHL} where it was
proposed that the imaginary part of the relaxation time can be
obtained from the dependence of the transition temperature $T_c$ on the
chemical potential $\mu$. Then  the first term in the GL thermodynamic
potential should be modified to
 $\Omega_{sc}=-\alpha (T_c+e\phi\partial T_c/\partial \mu
-T)\psi^{\ast}\psi+...$. Taking  into account that we should always
have the gauge invariant combination $(2e\phi-i\partial /\partial t)$
one obtains the imaginary part of the relaxation time in the TDGL
\cite{AHL,IKK} $\gamma_2=-{\alpha\over 2}\partial T_c/\partial\mu$.
Without Coulomb interaction the change in density can be obtained
in the same way as  before: $ n=-
\partial \Omega_{sc}/\partial\mu=const-2\gamma_2 |\psi(r)|^2$. Then
the $\delta n $ contribution to the Hall conductivity in Eq.
(9) coincides with the result of \cite{D,IKK} if the numerical
parameter $\beta$ ($-\alpha_2$  in notation of \cite{D})
is equal to 1,
which corresponds to the value of the TDGL parameter $u\ll 1$
\cite{IKK}.
For  large values of $u$ the analysis of \cite{D,IKK} gives a similar
result but with an additional coefficient of the order of unity in
front of the `$\delta n $' term. In these papers the condition
of the local electroneutrality  ($\nabla {\bf j}=0$, i.e. $\delta n(r)=0 $)
was imposed in order to take Coulomb screening into account.
Actually for the consistent treatment of the Coulomb effects
one should add a term $\phi^2/r_D^2$ to the GL free
energy and allow for local density variations.
The microscopic calculation for superconductors with paramagnetic
impurities \cite{LO} shows that these numerical corrections
to `$\delta n$' term become small for the low enough concentration
of paramagnetic impurities.

The effect of the vortex charge on the Hall effect was recently
considered by Khomskii and Freimuth \cite{KF}. Although the
treatment of the static charge distribution in the vortex core
is the same as ours, the transverse force and the Hall conductivity
found in \cite{KF} are much smaller (by a factor $\sim B/H_{c2}$) and
have {\it the opposite sign} as compared to our Eqs. (5,9), which
explicitely contradicts well-known result for the Magnus force in
the Galilean invariant case where $n_0=0$.

 The crucial
point for the discussion is the sign of $\delta\tilde n $. Taking as an
estimate $\delta\tilde n/n={\rm sign}(\delta\tilde n)(\Delta/E_F)^2$ and
$\omega_0=\Delta^2/E_F\ll \tau^{-1}$ one arrives at
\begin{equation}
\sigma_{xy}\simeq{n_0ec\over B}{\Delta^2\over E_F^2}((\Delta\tau)^2
-{\rm sign}(\delta n)).
\end{equation}
The new term we found  is important in the dirty case
$\Delta\tau<1$ and can lead to the sign change if $\delta\tilde n >0$ (the
 carrier density in the core is bigger than outside). Let us
consider this case in more details in application to HTSC. In this
materials $\Delta\tau>1$  at low temperature and
$\Delta\tau\rightarrow 0$  at $T_c$. Note that what enters in Eq. (5)
is $\Delta (T)$ rather than $\Delta(0)$. Thus at low temperatures
we can neglect this $\delta\tilde n$ contribution and the sign of
$\sigma_{xy}$ is positive (as in the normal state). As the temperature
approaches $T_c$,
 $\Delta\tau\sim 1$ and $\sigma_{xy}$ becomes  negative.
 Near $H_{c2}(T)$ $\omega_0 \approx$ cyclotron freqyuency $\omega_c$,
thus the first term in Eq.(11) transforms
to the normal state
Hall conductivity $\sigma^n_{xy}$, whereas the `$\delta n$'  contribution
 goes to zero  $\propto H_{c2}-H$, and the Hall effect changes
sign back to the normal value in this region \cite{IKK}. These are
just the two
sign changes observed in Bi and Tl based materials.
 In 90 K YBCO the low temperature sign change back to the
normal sign is usually not observed
 since  $\rho_{xy}$ is unmeasurably small because of  pinning.
However in the experiments where pinning was suppressed either
by a high current \cite{K} or by high frequencies \cite{berkeley} the
second sign change seems to be observed at low temperatures.
The temperature dependence of the Hall
conductivity (9) is in very good agreement with the data by Samoilov
{\it et al.} \cite{S} who found for TBCCO
that the $B^{-1}$- term in the Hall conductivity changes sign around 83 K
and at higher temperature is $\propto T_c-T$.

In the \cite{O1} the Hall angle evidence for the superclean regime
in 60 K YBCO was reported. In this material Hall angle changes sign
and becomes   relatively large ($\Theta_H\simeq -1/2$) at low temperature.
There are two quite different ways to treat these data in our scheme.
 The one taken in \cite{O1} is that in 60 K YBCO superclean limit is
realized with $ \omega_0\tau\gg 1$ and Magnus force has a `wrong'
sign due to the complicated structure of the Fermi surface. Another
possible scenario is that this material is not the superclean
but just moderately clean, with
$ \omega_0\tau\ll 1$,
 and has the same sign  of $\delta\tilde n$ as in 90 K material,
but with  larger numerical value (due to the fact that 60 K compound
is closer to the half filling and the dependence on
chemical potential  is sharper
than for the 90 K compound). In order to estimate the value of $\delta
\tilde n/n$  we note that the additional term in the Hall conductivity is
$-\delta\tilde n ec/B$, whereas the normal state Hall resistivity
is $\rho_{xy}^n=B/nec$. Multiplying these two quantities one can get
an estimate for $\delta\tilde n/n$. Analysis of the experiments
\cite{O1,S} gives $\delta\tilde n/n\simeq 10^{-3}$ for Tl compound
and 0.03 and 0.07 for 90 K and 60 K YBCO respectively.
Thus the difference between these two Y based materials seems to be
much smaller than between Y and Tl based compounds. Then the experimental data
\cite{O1} for the 60K
YBCO  can be understood under the assumption that at low temperatures
$\omega_0\tau\simeq 0.1\ll 1$, i.e. of the same order of
magnitude as $\delta\tilde n/n$; in that case
only the second term in (9) is important and the
Hall angle acquires the value of the order unity (since the logitudinal
conductivity
contains factor $\omega_0\tau$), although the material can still be
rather dirty
(note also, that 60 K material is traditionally considered as more dirty than
the 90 K one).  On the other hand, the 90 K YBCO is expected to have bigger
low-$T$ value of $\omega_0\tau$ and smaller (as estimated above) value of
$\delta\tilde n/n$, which makes the observation of the second sign change
in this material \cite{K,berkeley} quite natural.
The proposed second scenario seems preferable to us since it does not involve
any {\it ad hoc} hypothesis about the complicated Fermi-surface, and suggests
an unified decription of both 60 K and 90 K compounds.

In the simple BCS model $T_c$ depends upon the density of states and
increases  with increased density leading to
the positive $\partial T_c/\partial\mu$ and thus $\delta
n<0$. However one can consider a simple tight binding model with large
effective mass  exponentially dependent upon the lattice constant.
Then under compression carriers become lighter and $T_c$
decreases leading to  $\delta n>0$. The case of HTSC
is complicated by the fact that the normal state Hall effect
 has hole like sign, although from the simple electrons counting
 the Fermi surface should have an electron
like shape. It would be tempting to relate $\delta n$ term with
the doping dependence of $T_c$ via Eq. (11), which would lead to a conclusion
that the sign change should occur for the overdoped materials.
This is dangerous, however, since in some versions of the RVB-like
theories \cite {Lev} the doping dependence of $T_c$ and superconducting energy
away from $T_c$ may have opposite signs.

We thank Y. Imry and S. Levit for attracting our attention to
the Ref. \cite{AT} before its publication and B. Altshuler,
G. Blatter, L. Ioffe, N. Kopnin, A. Millis, A. van Otterlo and A. Samoilov
for very helpful discussions.  We acknowledge support from the
 Swiss National Foundation and the Landau Weizmann program for
Theoretical Physics. VMV and AIL acknowledge the support through US
Department of Energy BES-Materials Sciences, under contract
No. W-31-109-ENG-38 and MVF and VBG acknowledge the support from ISF
grant No. M6M000. The work was in part
supported by the ITP of the UCSB through Grant No. PHY89-04035.

\end{document}